\pdfoutput=1

\documentclass[sigconf]{acmart}
\copyrightyear{2020}
\acmYear{2020}
\setcopyright{licensedothergov}
\acmConference[I3D '20]{Symposium on Interactive 3D Graphics and Games}{May 5--7, 2020}{San Francisco, CA, USA}
\acmBooktitle{Symposium on Interactive 3D Graphics and Games (I3D '20), May 5--7, 2020, San Francisco, CA, USA}
\acmPrice{15.00}
\acmDOI{10.1145/3384382.3384522}
\acmISBN{978-1-4503-7589-4/20/05}


\usepackage[T1]{fontenc}

\usepackage[utf8]{inputenc} 
\usepackage[T1]{fontenc}    

\usepackage{xcolor}
\usepackage{graphicx}
\usepackage{subfigure}
\usepackage{wrapfig}
\usepackage{booktabs}       
\usepackage{amssymb}
\usepackage{amsmath}
\usepackage{amsfonts}       
\usepackage{microtype,balance,stfloats}      

\newcommand{\bandproc}{$\mathcal{B}$}
\renewcommand{\paragraph}[1]{\noindent \textit{#1}}
\newcommand{\mfx}{Universit\'{e} de Lorraine, CNRS, Inria, LORIA}
\renewcommand{\shortauthors}{J. Etienne, S. Lefebvre}

\begin{document}

\title {Procedural band patterns}
\author{Jimmy Etienne}
\affiliation{ \institution{\mfx}}
\email{jimmy.etienne@inria.fr}
\author{Sylvain Lefebvre}
\affiliation{\institution{\mfx}}
\email{sylvain.lefebvre@inria.fr}

\begin{abstract}
We seek to cover a parametric domain with a set of evenly spaced bands which number and width varies according to a density field. 
%
We propose an implicit procedural algorithm, that generates the band pattern from a pixel shader and adapts to changes to the control fields in real time. 
Each band is uniquely identified by an integer. This allows a wide range of texturing effects, including specifying a different appearance in each individual bands. Our technique also affords for progressive gradations of scales, avoiding the abrupt doubling of the number of lines of typical subdivision approaches.
This leads to a general approach for drawing bands, drawing splitting and merging curves, and drawing evenly spaced streamlines. Using these base ingredients, we demonstrate a wide variety of texturing effects.
\end{abstract}


\begin{teaserfigure} 
  \centering
  \includegraphics[trim={0cm 0cm 0 0cm}, clip, width=\textwidth]{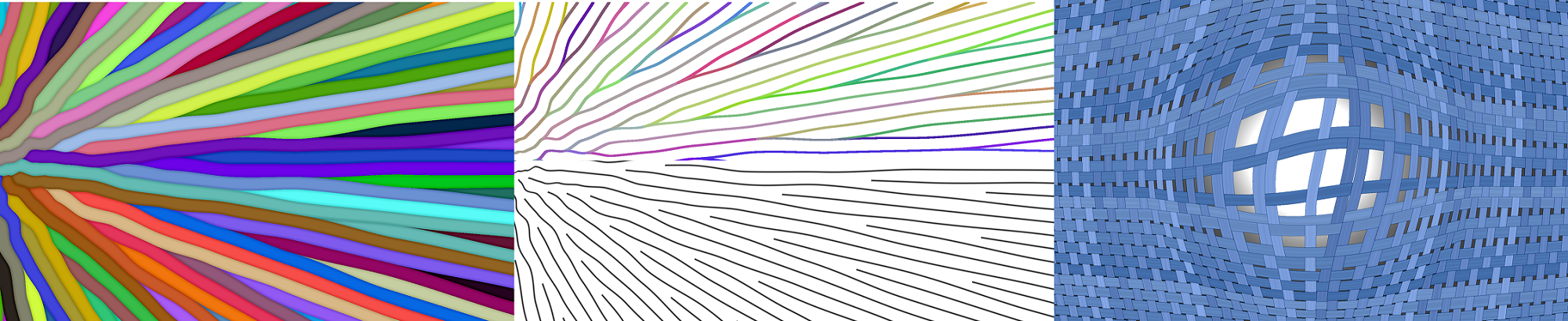}
  \caption{
  Our technique generates band patterns following a parametric field, while adapting the density of bands per unit. Each band is uniquely identified (\emph{left}) which affords for robust extraction of border trajectories and center lines (\emph{middle}). The method is procedural which allows a wide range of dynamic shader effects, such as textile patterns that adapts to stretch (\emph{right}). Contrary to classical subdivision approaches, our approach introduces new bands with a non power of two factor, allowing a more progressive gradation.
  }
  \label{fig:teaser}
\end{teaserfigure}


\maketitle

\section{Introduction}

We focus on producing patterns of parallel bands along a given parametric field, while locally adapting the number of bands to an input control field.
More precisely, we expect as input a domain $\Omega \in \mathbb{R}^{2}$ or $\Omega \in \mathbb{R}^{3}$, a parameter field $u : \Omega \rightarrow \mathbb{R}$, and a density field $d : \Omega \rightarrow \mathbb{R}$ that defines a target number of bands per unit of $u$.
These fields may be the direct result of an optimization~ \cite{groen2019homogenization} or may be painted by the user. The field $d$ may be linked to $u$ (e.g. compensating for a distortion) or may be independent from it, in which case the actual local density of bands will be a combination of the intrinsic stretch of $u$ and the density factor of $d$.
Both $u$ and $d$ are assumed to be continuous, smooth scalar fields.

We denote by $p$ a point in $\Omega$, such that we obtain the parameter value $u(p)$ and density value $d(p)$ at each point $p$.
The objective is to produce a set of bands which are flowing along the isovalues of $u$ and where the local density of bands -- or equivalently their spacing -- is changing according to $d$.

The originality of our approach is to define a lookup function
\bandproc{}~$(v,s) : \mathbb{R} \times \mathbb{R} \rightarrow \mathbb{N}$ 
that returns the unique integer ID identifying the band enclosing parameter value $v$, and where the local density is controlled by $s$. In our context the lookup is performed for a point $p \in \Omega$ as \bandproc{} $(u(p),d(p))$. 
\bandproc{} is \textit{procedural}~\cite{lagae:2010}: it has a minimal computational and memory requirement, allowing its implementation in a pixel shader for fast synthesis and interactive manipulation and animation of complex patterns. 
The returned integers uniquely identify the bands allowing for a wide range of texturing effects.


\paragraph{Previous work}
Drawing evenly spaced curves and streamlines is a long standing problem in Computer Graphics. We identify three main families: approaches tracing streamlines in a vector field from starting points~\cite{Jobard:97,Spencer:2009,Mebarki:2005,Hertzmann:2000}, approaches based on global periodic parameterizations~\cite{stripes} and finally approaches splitting and merging covering curves, as pioneered by the seminal work of Elber and Cohen~\cite{Elber96}. Our approach most closely relates to this later family, and in particular its image space variants~\cite{groen2019homogenization}.

\paragraph{Contributions.}
Our approach introduces several improvements. First, by defining numbered bands as opposed to directly curves, we produce a partition of the domain $\Omega$, enabling a robust (and simple!) extraction of each region and its boundary. The unique identifier also affords for direct manipulation of the bands and their borders, for texturing and patterning effects. For instance, drawing the center line of fully deployed bands produces evenly spaced streamlines.
Second, existing subdivision techniques define gradation by doubling the density. This leads to abrupt jumps in the number of curves and provides only a crude approximation of the target scale. Instead, our technique allows a finer control. 
Finally, our approach is fully implicit and defined by a procedural lookup function, avoiding global geometric constructions and optimization. We are not aware of any other technique offering a similar capability.


%



\section{Method}


Given the fields $d$ and $u$, we cover the domain $\Omega$ with a discrete grid and synthesize the bands by evaluating \bandproc ~ on every node. This is implemented as a pixel shader (GLSL) applied to a render target. The code is given in Section~\ref{sec:code}. The remainder of the text progressively introduces the principles and elements of the algorithm.

\subsection{Overview}

Given a spacing $s$, it is trivial to produce parallel bands in $u(p)$ by defining the band identifier as $\lfloor \frac{u(p)}{s} \rfloor$. The principle of our approach is to cover the domain with many overlapping sets of parallel bands, using different spacings. The spacings are decreasing powers of a base value $1 < step \leq 2$. 
We call \textit{bands of level} $L$ the set using a spacing of $step^{-L}$.
The choice of $step$ allows a trade-off between a doubling split ($step = 2$) or a more progressive gradation, which we discuss in Section~\ref{sec:controls}. 

We select which set should appear in a given location following $d(p)$, and interpolate (deform) the bands, with smaller bands borders joining or closing onto coarser band borders. This in turn defines a parent--child relationship which we use to define a global numbering through the band hierarchy.

\subsection{Procedural bands}

Given a point $p$, we seek to compute the global identifier of the band enclosing $p$. This is done in two steps. First we identify the {local ID} of the band to which $p$ belongs at the density $d(p)$. The local ID is the rank of the band in its set, it is computed by function \texttt{getLocalID} (see code Section~\ref{sec:code}). Second, we produce a {global ID} for the band in function \texttt{getGlobalID}.

\begin{figure}[tb]
\centering
\includegraphics[width=0.8\linewidth]{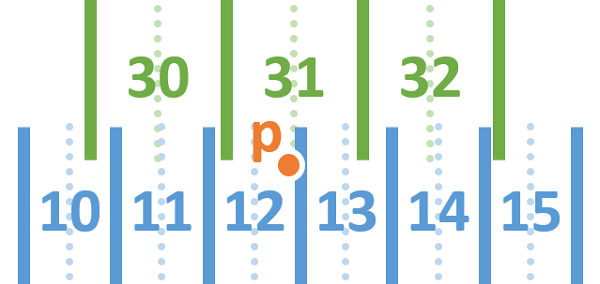}

\caption{\label{fig:quantize}Every point $p$ belongs to a band in a finer and a coarser set enclosing the target density $d(p)$: the finer set has more bands than desired at $d(p)$ while the coarser set has fewer bands.}
\end{figure}

We now focus on \texttt{getLocalID} (line 5). We first determine from $d(p)$ the band levels bracketing the target density. This is achieved by quantizing $d(p)$ based on the value of $step$, see the \texttt{quantize} function line 1. 
This gives the situation illustrated in Figure~\ref{fig:quantize}: $p$ belongs to two bands, one in a set having more bands than the target $d(p)$, and one in a set having fewer bands.

\begin{figure}[tb]
\centering
\includegraphics[width=0.8\linewidth]{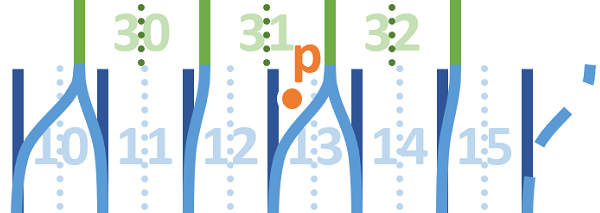}

\caption{\label{fig:interp}The borders of the finer bands are pulled towards their closest coarser band border. A point $p$ in a transition area may end up in a different band after interpolation.} 
\end{figure}

We then define an interpolation that displaces the band borders, making the finer bands deform towards their coarser counterparts. This is illustrated in Figure~\ref{fig:interp}. The interpolation is performed by pulling each finer band border towards its closest coarser band border. The interpolator is given by the position of $d(p)$ within the density interval of the bracketing levels.
Note that as we change the location of the band boundaries, $p$ may no longer be in the same initial band. This is tested and adjusted for lines 23 and 24.
Through interpolation, some finer bands are \textit{closed}: both their borders move towards the same coarser band border, e.g. band 13 in Figure~\ref{fig:interp}.



\begin{figure}[tb]
  \centering
  \includegraphics[width=0.8\linewidth]{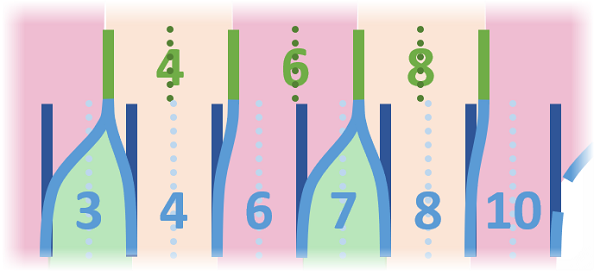}
  \caption{Band global IDs are computed in a hierarchical fashion, with parent defining most significant bits while children use least significant ones. The ID of closing bands no longer appears in lower levels, leaving a gap between IDs.}
  \label{fig:ids}
\end{figure}

After this process, we obtain the ID of the band enclosing $p$ in the level just finer than $d(p)$. This ID is local to this particular level: if used directly, it would change arbitrarily where a change of level occurs in $d$. Instead, we produce a global ID, valid \textit{across} levels.
This is achieved in \texttt{getGlobalID} (line 28). Note that since we have $step \leq 2$, we at most double the number of bands from one level to the next. Thus, our approach is to number the bands in the same way we number leaves in a binary tree. From the current band level, we move up the hierarchy (lines 32--52), and insert a bit in the ID (line 43) each time the band closes (lines 35--41). 
Finally, the ID of the top level band is added as the most significant bits of the global ID (line~54). This is illustrated in Figure~\ref{fig:ids}. 
In other words, the global ID is the ID of the top level band, followed by the binary pattern of opening bands to reach the band enclosing $p$.

As an added bonus, we can easily detect when a band \textit{just appears}, by checking if the first band is closing immediately (lines 44--47). This is an interesting capability, as such a band is one that is smaller than the target spacing -- there is not yet enough space between the parents for it to be fully deployed. In particular, we may choose to remove such a band or draw it in a different style.


\begin{figure}[tb]
  \centering
  \includegraphics[trim={0cm 0cm 0 0cm},clip,width=1.0\linewidth]{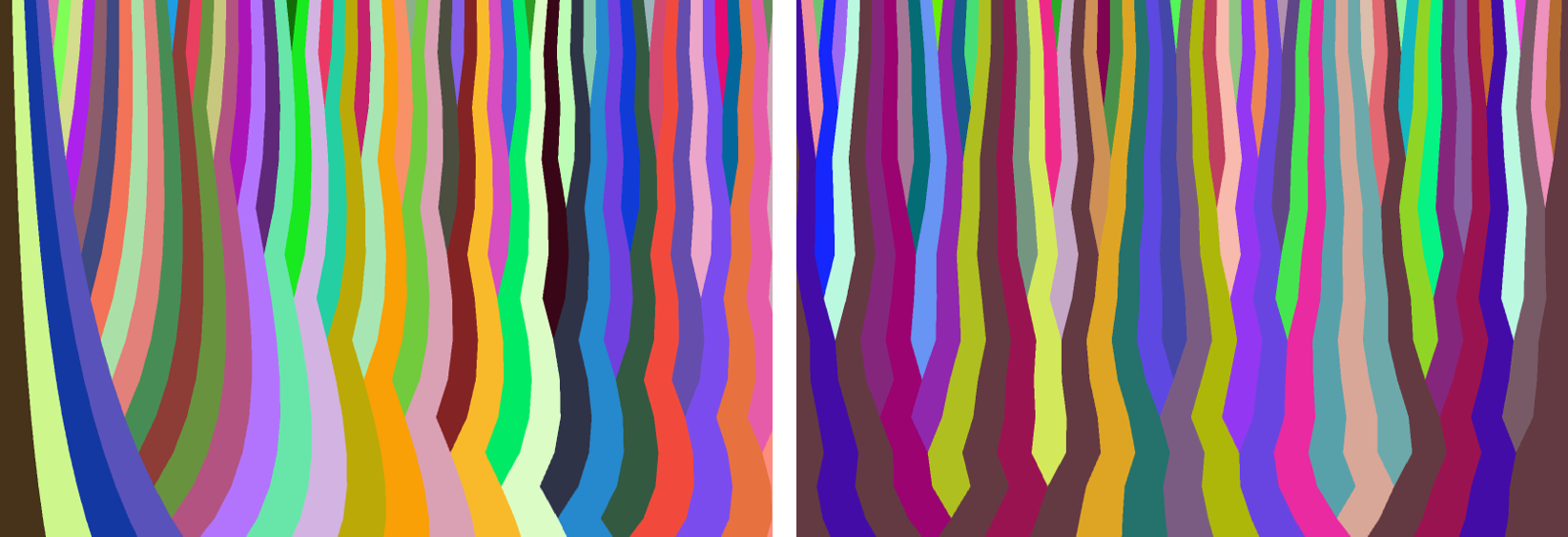}
  \caption{
  \emph{Left:} Without shifts, the alignment of all band sets at the origin produce an artificial, peculiar pattern. \emph{Right:} The random shifts break the alignment, making the pattern similar everywhere.}
  \label{fig:shift}
\end{figure}

\paragraph{Band shifting} Using an irregular subdivision pattern ($step<2$) poses an interesting challenge. Around the origin, all band sets align, producing an undesirable pattern shown in Figure~\ref{fig:shift}, left. This can be suppressed by translating the band sets by a (pseudo-)random shift, which is different for each band level. The effect is shown in Figure~\ref{fig:shift}, right.

For $step=2$, we shift the bands at every level by half their spacing to obtain the traditional balanced subdivision pattern.

\section{Controls and parameters}
\label{sec:controls}

\paragraph{Progressive increase in density}
By changing $step$, we control the number of bands opening at every level. For $step=2$ our technique behaves similarly to traditional subdivision approaches, with one new band opening for each coarser band, effectively doubling the number of bands at every level.

It is worth outlining that this approach, while very typical, also gives only a crude approximation of the target density. Indeed, across an interpolation range the produced density (number of bands) is wrong by as much as 50\% (in the middle).

Using lower values of $step$, our technique affords for a very progressive insertion of bands at every new levels. Let us consider a value $step = \frac{N}{M}$ with $M,N \in \mathbb{N}$ forming an irreducible fraction, and $M < N \leq 2M$. Such a step will open $N-M$ new bands for every $M$ bands in the previous level -- this stems from the fact that each level $L$ corresponds to a density $step^L$.

$N=2, ~ M=1$ gives the standard case (100\% increase every level). $N=3, ~ M=2$ opens one new band every two coarser bands (50\% increase). $N=17, ~ M=13$ opens four new bands every thirteen coarser ones (31\% increase). These number can be freely chosen as long as $step$ remains in the $]1,2]$ interval.

\paragraph{Regularity and periodicity}
The produced pattern is entirely defined by the subdivision of the top level bands (for $d = step$, we have one band per unit in $u$). It is intriguing to consider the periodicity of the split pattern. For the sake of clarity we ignore the random shifts in this discussion and assume all band sets are aligned on the origin.

Let us again write $step$ as an irreducible fraction $\frac{N}{M}$. From one level to the next, a period occurs in the split pattern every $M$ bands, as borders across both levels align: this repeats the pattern at the origin. However, this is the case for \textit{two consecutive} levels. When we consider more levels, the period before all borders align grows. In fact, for $k$ levels the period becomes $M^k$. Indeed, given $Q$ bands at the base level, the number of bands at level $k$ is $Q\left(\frac{N}{M}\right)^k$. This number will only be an integer value for $Q = M^k$, since $M$ and $N$ are coprime (irreducible fraction). 







\paragraph{Interpolation profile}
All our results use a simple linear interpolation. However, using a different profile (e.g. GLSL smoothstep) leads to different (smoother) band shapes in the opening region. In addition, the interpolation profile can control how 'quickly' new bands open to their full width.


\section{Pseudo-code}
\label{sec:code}

The detailed pseudo-code is given next. We also refer the reader to our shadertoy implementations (see Table~\ref{tbl:links}).

\noindent\includegraphics[width=.38\textwidth]{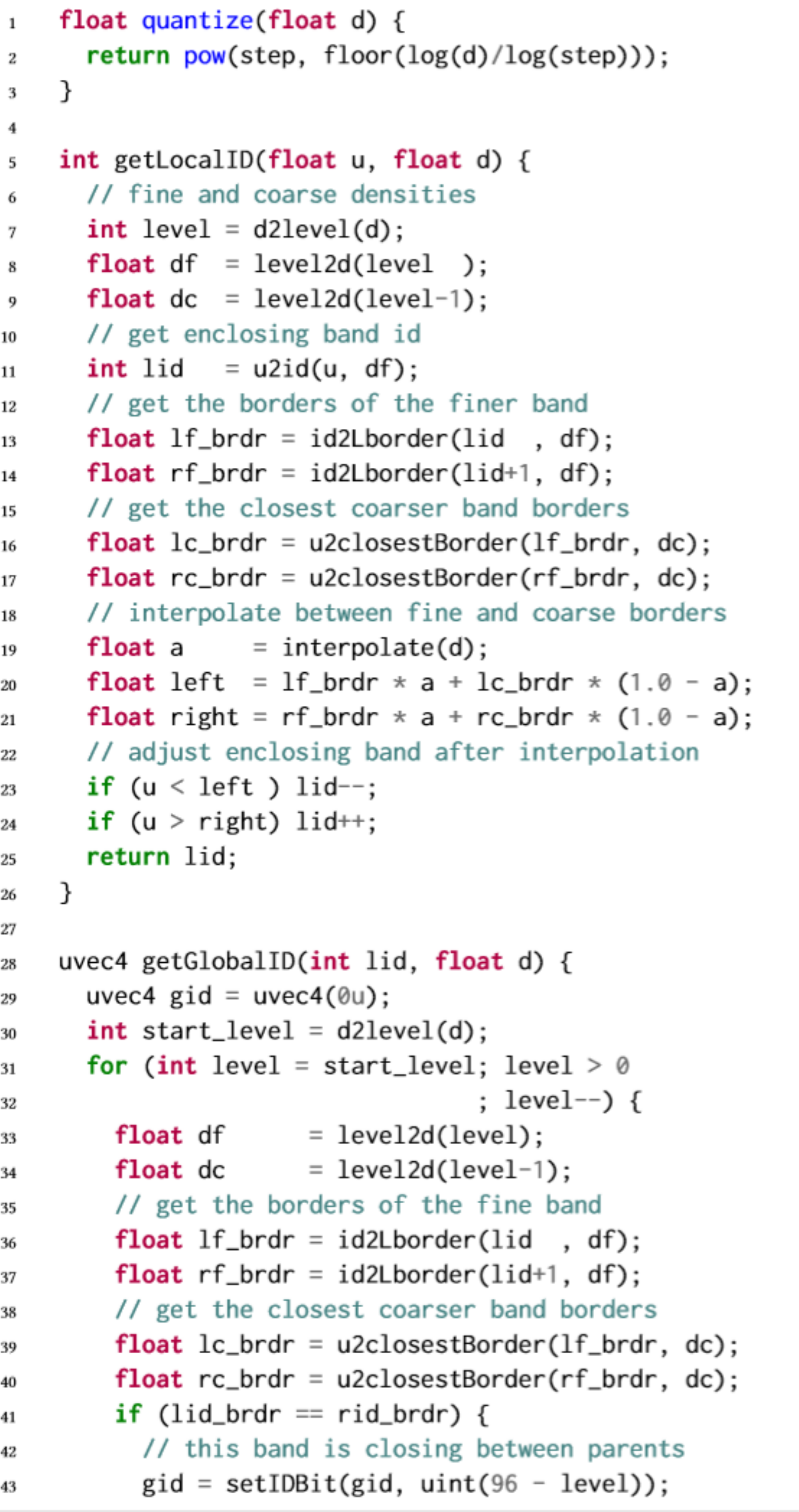} 

\noindent\includegraphics[width=.38\textwidth]{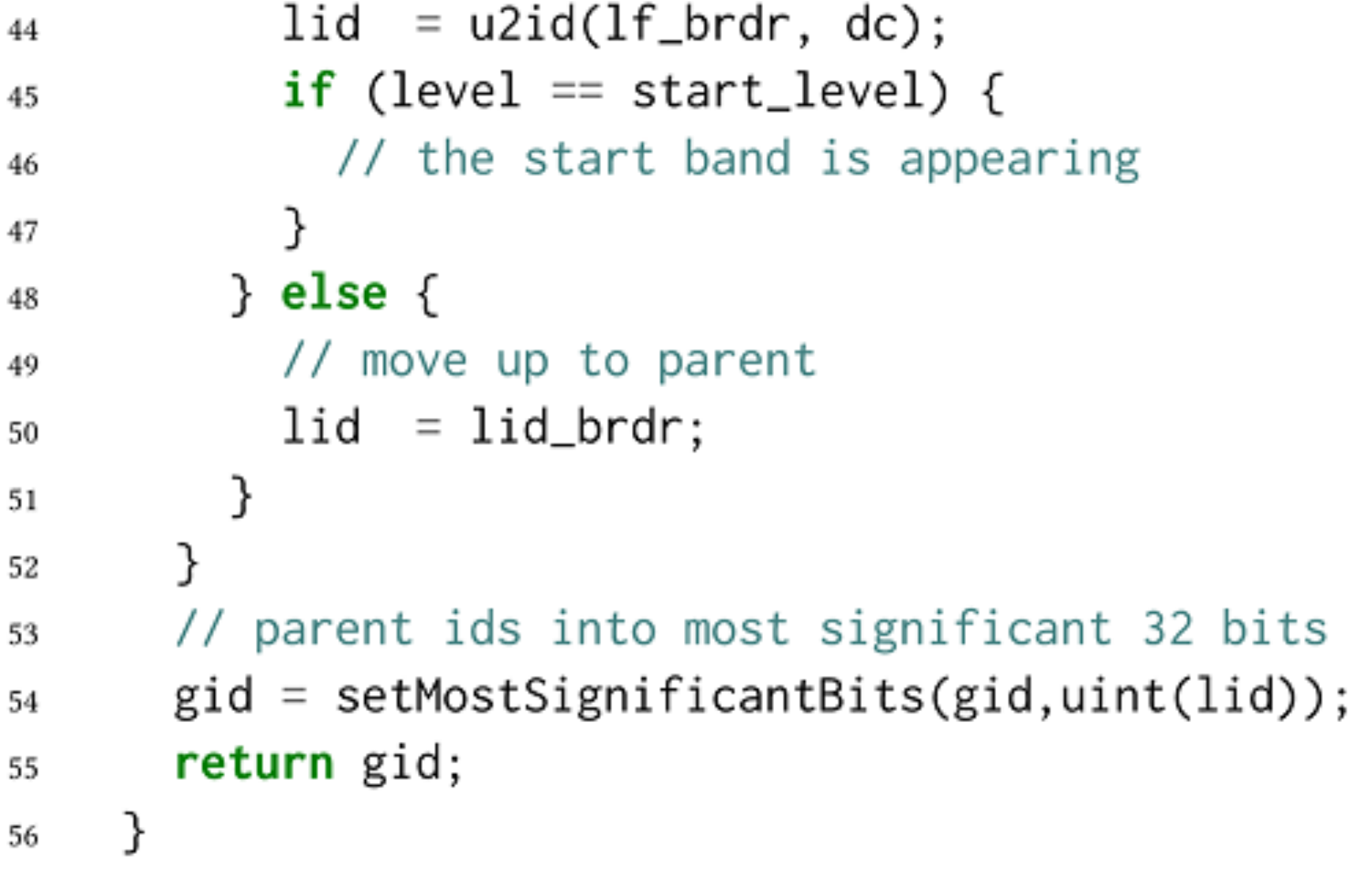}

\section{Results}

Our technique opens multiple possibilities to produce band and curve patterns. In this section we first show some basic patterns obtained directly from our technique, and discuss the effect of parameters in Section~\ref{sec:res:basic}. We then discuss how to use various properties to produce more elaborate animated patterns in Section~\ref{sec:res:anim}. Finally, we discuss how trajectory extraction can be used to produce graded fill patterns for additive manufacturing in Section~\ref{sec:res:infill}.

\subsection{Bands and curves}
\label{sec:res:basic}

\begin{figure*}[tb]
  \centering
  \subfigure[Radial]{
    \begin{minipage}[c]{0.14\linewidth} 
    \includegraphics[width=\linewidth]{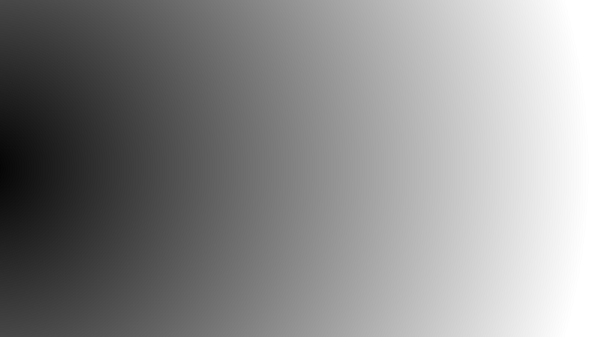} 
    \includegraphics[width=\linewidth]{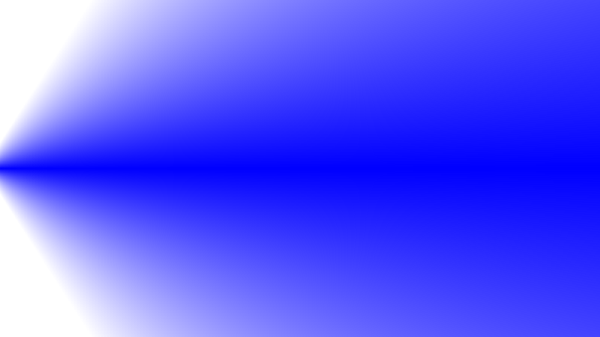}
    \end{minipage}
    \begin{minipage}[c]{0.85\linewidth} 
    \includegraphics[width=0.33\linewidth]{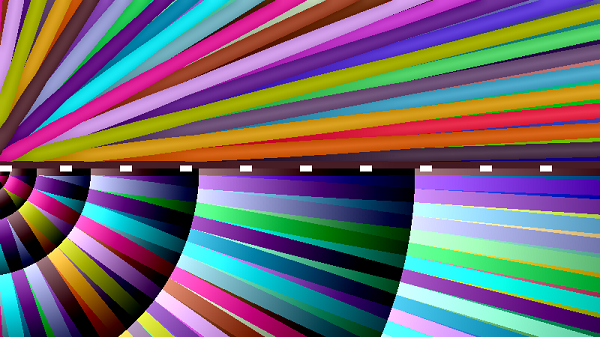}
    \includegraphics[width=0.33\linewidth]{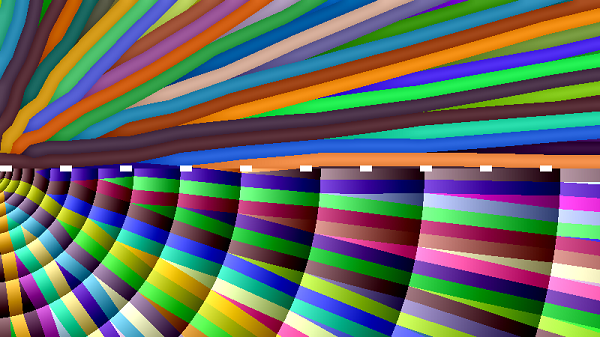}
    \includegraphics[width=0.33\linewidth]{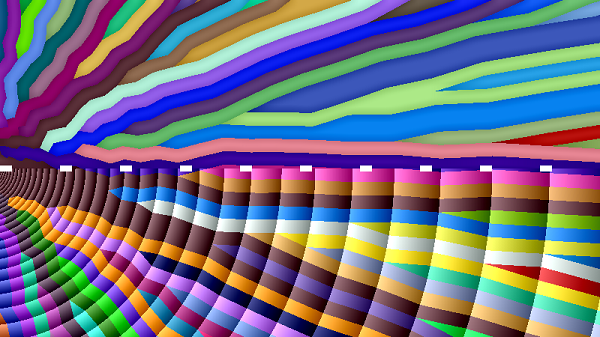}
    \end{minipage}
  }
    \subfigure[Swirl]{
    \begin{minipage}[c]{0.14\linewidth} 
    \includegraphics[width=\linewidth]{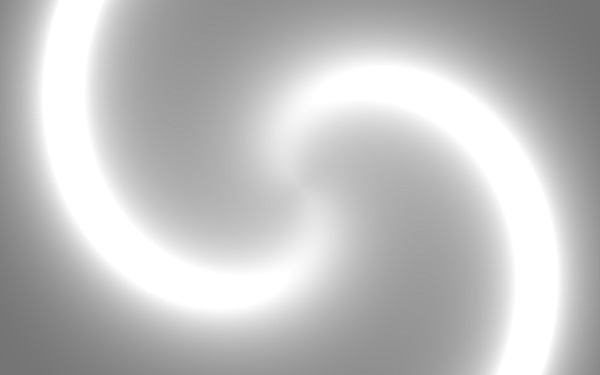} 
    \includegraphics[width=\linewidth]{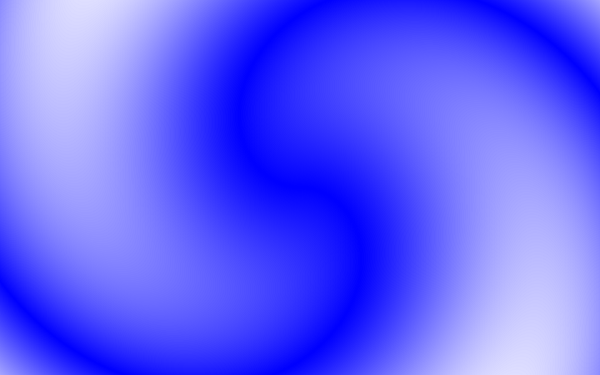}
    \end{minipage}
    \begin{minipage}[c]{0.85\linewidth} 
    \includegraphics[width=0.33\linewidth]{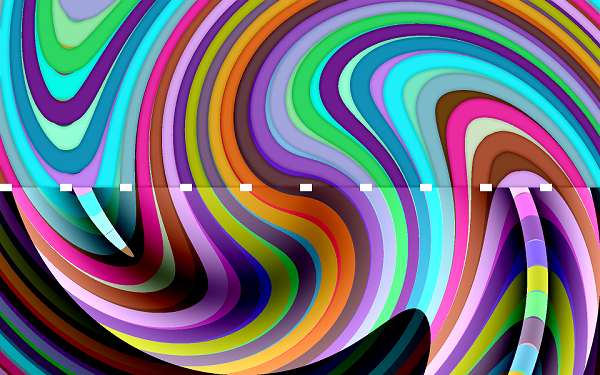}
    \includegraphics[width=0.33\linewidth]{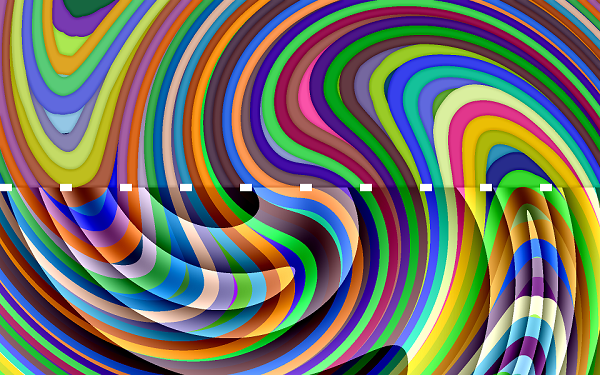}
    \includegraphics[width=0.33\linewidth]{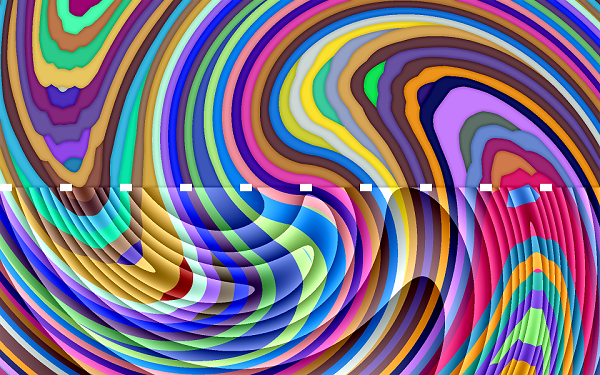}
    \end{minipage}
  }
  \caption{Two results using complex $u$ and $d$ fields, using $step=2$, $\frac{17}{13}$, $\frac{79}{71}$. \textbf{Left:} $u$ field in blue-white, $d$ in shades of gray. \textbf{Below dashed line:} discretization steps of $d$ overlaid, the more steps the more precise is the fit between the actual and the target number of bands.
  }
  \label{fig:results}
\end{figure*}

Figure~\ref{fig:results} shows two complex cases of fields $u$ and $d$ with bands produced for various values of $step$. The left most is $step=2$ while the others are showing decreasing values ($step=\frac{17}{13}, step=\frac{79}{71}$).
The parts below the dashed lines are showing the interpolation regions (discretization of $d$). Note how as $step$ decreases, the discretization of $d$ refines, producing a better match to the input field. For small values of $step$, the distribution of splits becomes less regular but the bands are also more 'jaggy'. A suitable tradeoff is easily found by interactive manipulation.

Figure~\ref{fig:teaser} shows three drawing modes enabled by our approach. 
The first is to extract a network of splitting/merging curves from the boundaries of the band regions. Here, the band ids allow to robustly decide which pixel edges belong to a curve, and also provide an ID for each extracted trajectory (the pair of band IDs on either side). 
The second is to draw the middle lines of fully deployed bands (see test lines~44--47) to obtain evenly spaces streamlines. 
The third is to draw bands directly, here with a subtle shadowing effect. Much more is possible in terms of patterning, as we will discuss next.

\subsection{Animated patterns}
\label{sec:res:anim}

\noindent \textit{Note: These effects are best seen animated. We provide the shadertoy links at the end of this section (Table~\ref{tbl:links}).}
 
We provide examples of animated effects, exploiting several unique possibilities. We generate the effects by having procedural fields $u$ and $d$ dynamically drive one or multiple band patterns. The $u$ field is obtained analytically (sum of sines, radial basis functions and noise), while $d$ is often computed from the gradient of $u$ to compensate for distortion. We then use band IDs as well as the ability to compute a local parameterization (distance of lookup point to enclosing band left border) to achieve various effects. 

Simple yet interesting effects are obtained by coloring the bands in a stretching field with distortion compensation, see for instance \textit{flow bands}. Driving the band density also produces interesting results, such as in \textit{scaling bands} where the band density is increased in a moving circular area.
The \textit{band flag} effect combines two sets of adapting bands, with a coloring that retains consistency along each direction, producing a typical towel coloring scheme.

We also explored tearing animations, where we discard all lines appearing in between the main parent lines. This is easily done by checking the least significant bits of the band IDs. This produces an effect where the main bands maintain their width, while spaces appear in between. The parameter $step$ then controls how and whether the bands form groups under stretch. The shaders \textit{tear twist, lens, light claws, textile} and \textit{net} are obtained in this manner. 
Instead of discarding, we can also change the coloring of the bands, producing effects such as the \textit{hairdryer} shader. Here, new darker hair strands appear as the motion expands the textured area.

The combination of two sets of adapting bands and tearing leads to effect producing squares, for instance in \textit{lens} and \textit{light claws}.
The \textit{lens} effect is a lens where instead of zooming the squares maintain their size and the bands tear to accommodate for the distortion. Note the uneven split pattern thanks to a choice of $step < 2$.
The \textit{light claws} effect drives a distortion field from sound, producing dynamically splitting bands along lines in two directions, disconnecting squares.

We can also weave two orthogonal sets of bands to produce textiles, for instance in \textit{net} and \textit{textile}. We determine which band is front at crossings using binary checks on their IDs. 
We also the ability to compute a local parameterization within the bands width to make then thinner.
Note how under distortion the weaving expands and slides while the bands maintain their width, revealing the background.

All these shaders are dynamic and real time -- thus they would work in a rendering context, for instance to adapt a texture to stretch. As our technique only requires a single input coordinate ($u$), it can also be applied to solids and implicit surfaces.

{
\begin{table}
\caption{\label{tbl:links}Links to anonymized shaders.}
\small
\begin{tabular}{ll}
Flow bands    & \url{https://www.shadertoy.com/view/wlt3DM}\\
Scaling bands & \url{https://www.shadertoy.com/view/tlt3DM}\\
Tear twist    & \url{https://www.shadertoy.com/view/WttGD8}\\
Band flag     & \url{https://www.shadertoy.com/view/3ttGD8}\\
Textile       & \url{https://www.shadertoy.com/view/wl3GDH}\\
Net           & \url{https://www.shadertoy.com/view/3ldGD8}\\
Lens          & \url{https://www.shadertoy.com/view/WltGD8}\\
Light claws   & \url{https://www.shadertoy.com/view/tt33W7}\\
Hairdryer     & \url{https://www.shadertoy.com/view/3tdGW8}\\
\end{tabular}

\end{table}
}


\subsection{Infill patterns for 3D printing} 
\label{sec:res:infill}

We use our technique to produce fill patterns inspired from so-called cubic infills \cite{slblog,wu2016self}. We extract the trajectories as the contours between pixels associated with different band IDs. This process is fast and robust -- akin to extracting the contours of a white region in a binary image. The key advantage is that our technique allows for spatial grading of density while producing very continuous paths. This has been implemented (and shipping to users) in our slicer \cite{anonymous2} for more than two years, and was in fact the initial motivation behind this work.


There are other contexts in additive manufacturing where isolines of equal spacing are desirable, for instance to generate fill patterns following optimized fields~\cite{Steuben:2016} or for curved 3D printing~\cite{Ezair:2018}. We believe our approach to be especially promising to decompose a 3D shape into solid slabs, before filling them with curved paths (using contouring and zigzag fill within the extracted curved slab~\cite{Chakraborty2008EPG}).

\begin{figure}[tb]
  \centering
  \includegraphics[width=1.0\linewidth]{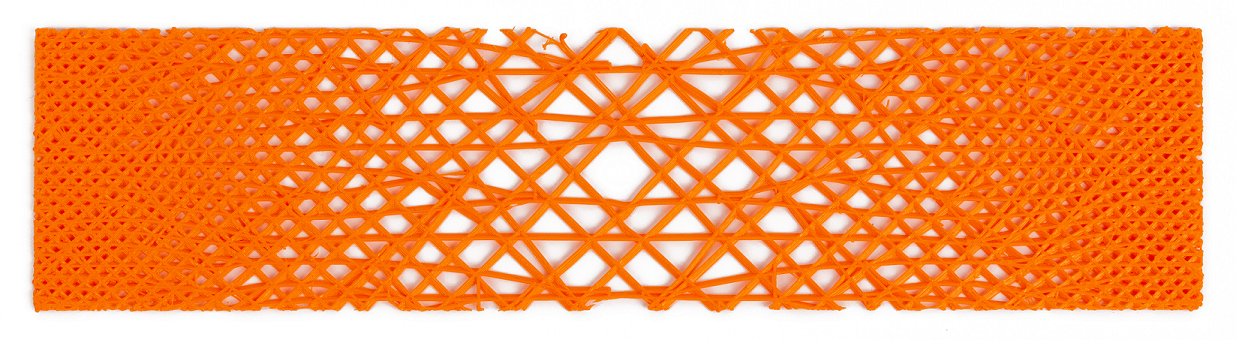}
  \caption{3D printed part revealing an infill pattern obtained from our technique. Note the very progressive change in density and the curve network extracted at each layer.}
  \label{fig:infill}
\end{figure}


\section{Limitations}


Our technique has a number of limitations, that can be problematic depending on the intended use.

In areas where the density field $d$ varies quickly, bands will appear distorted. This is for instance visible in Figure~\ref{fig:results}, bottom right, in areas of high curvature. It would be interesting to apply filtering to $d$ such as to limit such defects.

Intermediate density levels always exhibit partially deployed bands. A consequence is that, while the number of bands remains correct, their spacing may look uneven in areas of constant scale (linear $u$ and constant $d$). This situation may be minimized -- albeit not entirely resolved -- by modifying the interpolation profile to quickly open/close bands (see Section~\ref{sec:controls}). 

In general we believe it would be desirable to randomize the locations where the band splits occur. While our irregular splitting strategy ($step < 2$) reduces split alignments, they remain located along isolines of $d$ and not randomly distributed.

The numbering scheme (\texttt{getGlobalID}) relies on integers and might run out of numbers, being unable to distinguish between bands deep in the hierarchy. Using wider integers is not too difficult in our context as the only operation is to set bits in the IDs. Nevertheless infinite zooming would require recycling IDs.


\section{Conclusion}

We propose a fully procedural technique to generate band patterns. It is implemented as a fast pixel shader that interactively reacts to changes to the control fields. We believe our technique to be useful for a wide range of applications, such as visualization, texturing, but also for producing fill patterns in additive manufacturing. 

Our technique extends trivially to 3D, for instance allowing to decompose a shape into solid slabs -- the equivalent of our bands -- while retaining all the procedural and adaptive capabilities of our technique. We envision direct applications in modeling of solid properties and process planing for additive manufacturing.


\section{ACKNOWLEDGEMENTS}

The work was partly supported by Région Grand-Est, Lorraine Université d’Excellence (ANR-15-IDEX-04-LUE) and CNRS. We thank Cédric Zanni for proof reading.


\begin{figure*}
  \centering
  \includegraphics[clip,width=0.4\linewidth,height=4cm]{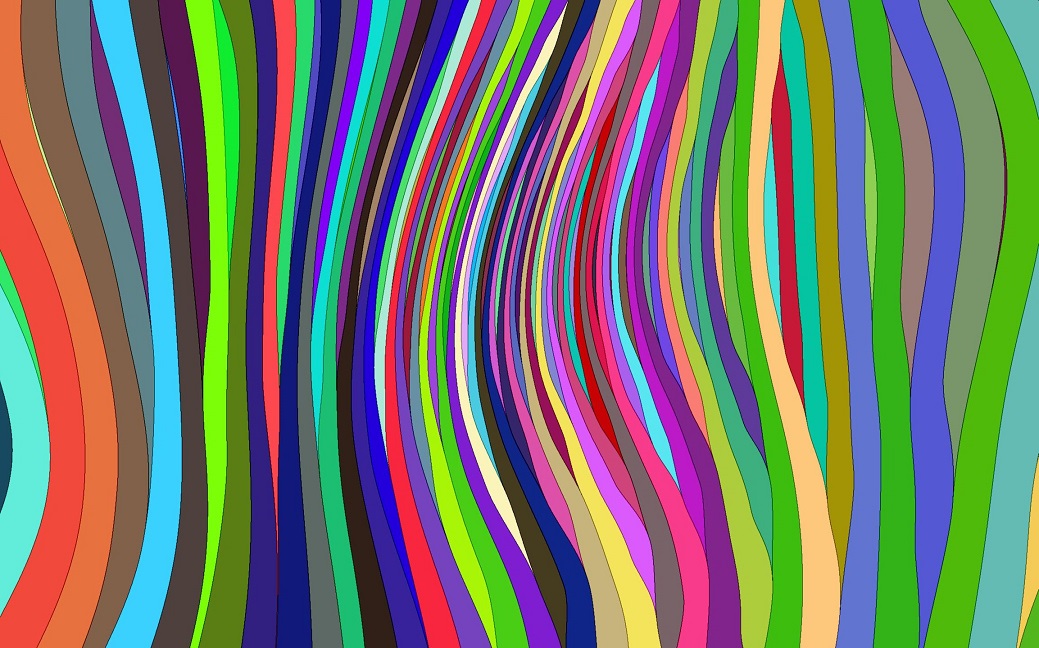}
  \includegraphics[clip,width=0.4\linewidth,height=4cm]{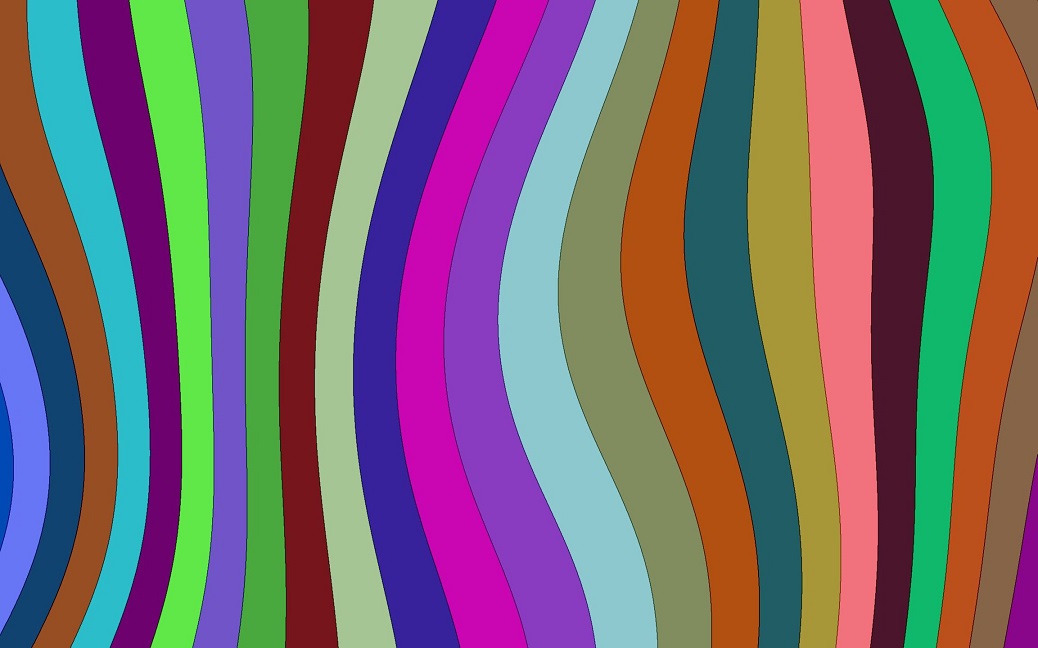} 
  \\
  \includegraphics[clip,width=0.4\linewidth,height=4cm]{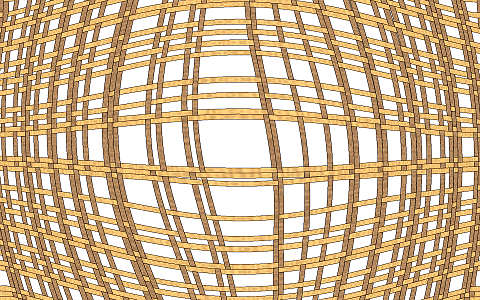}
  \includegraphics[clip,width=0.4\linewidth,height=4cm]{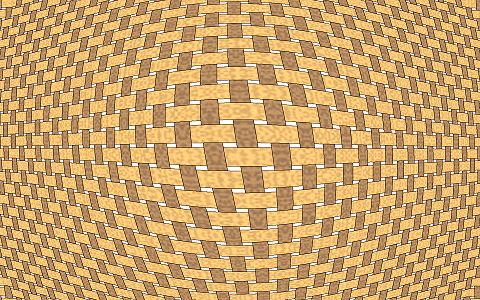}
  \\
  \includegraphics[clip,width=0.4\linewidth,height=4cm]{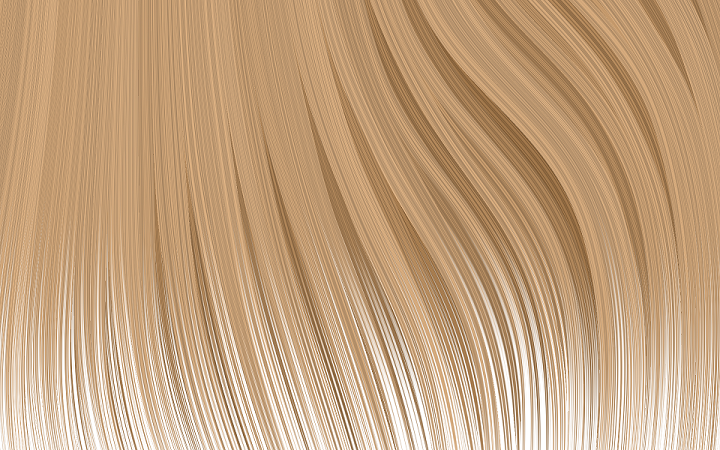}
  \includegraphics[clip,width=0.4\linewidth,height=4cm]{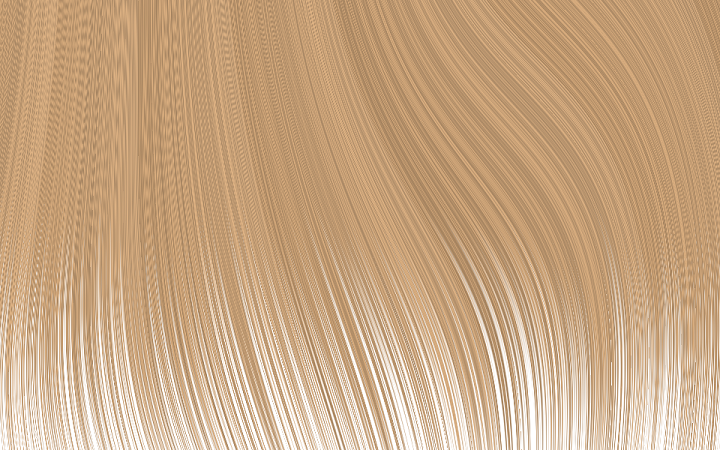}
  \\
  \includegraphics[clip,width=0.4\linewidth,height=4cm]{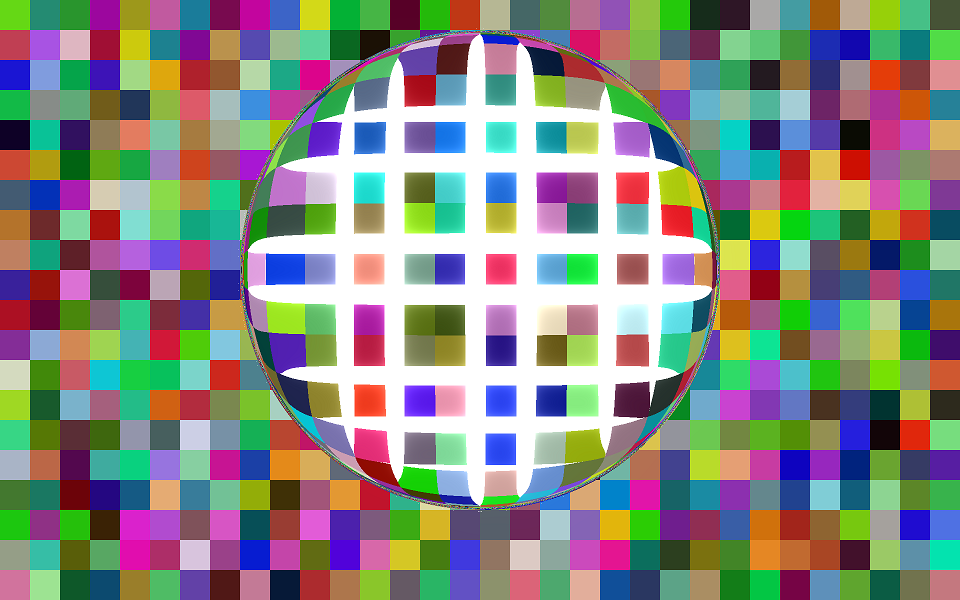}
  \includegraphics[clip,width=0.4\linewidth,height=4cm]{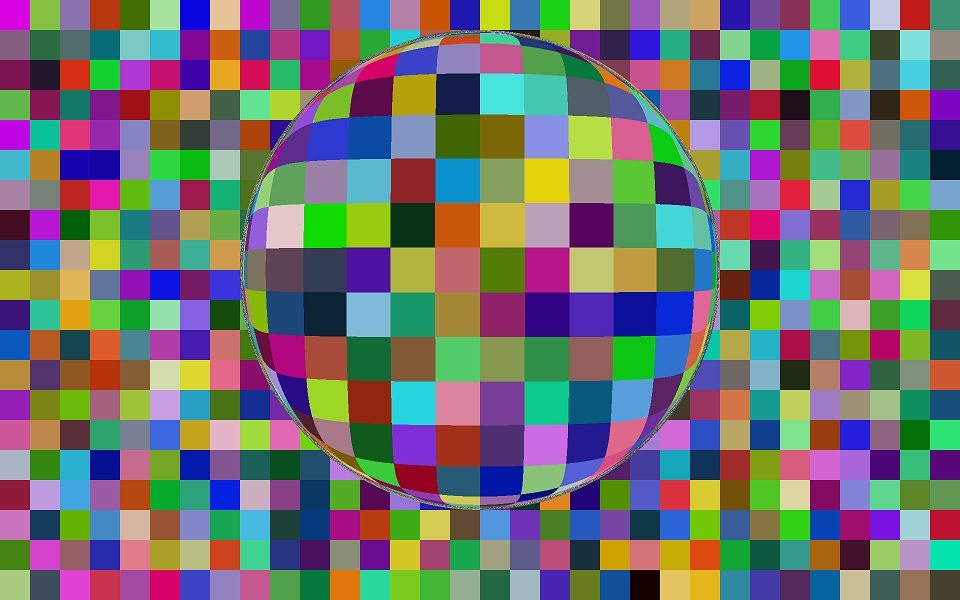}
  \\
  \includegraphics[clip,width=0.4\linewidth,height=4cm]{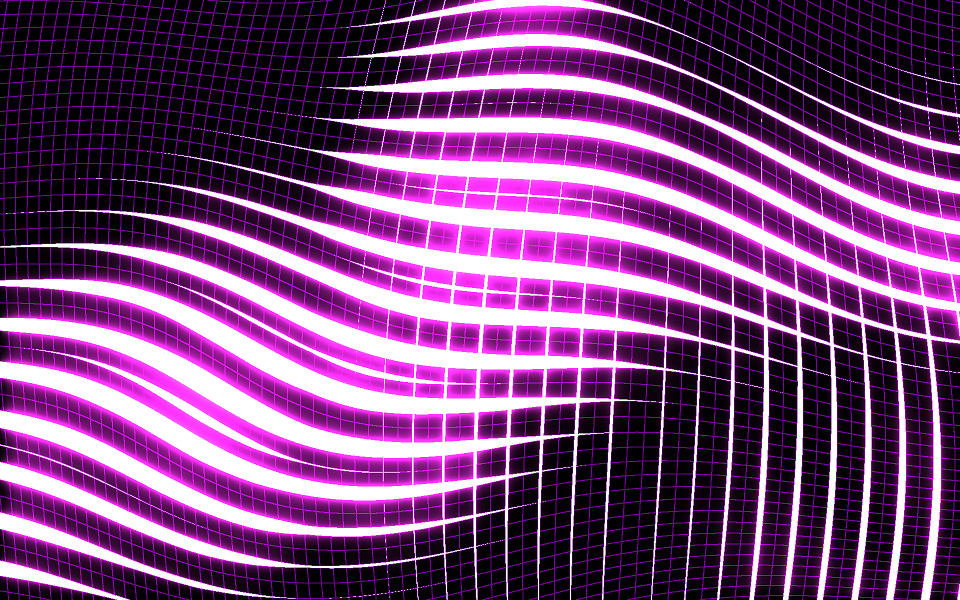}
  \includegraphics[clip,width=0.4\linewidth,height=4cm]{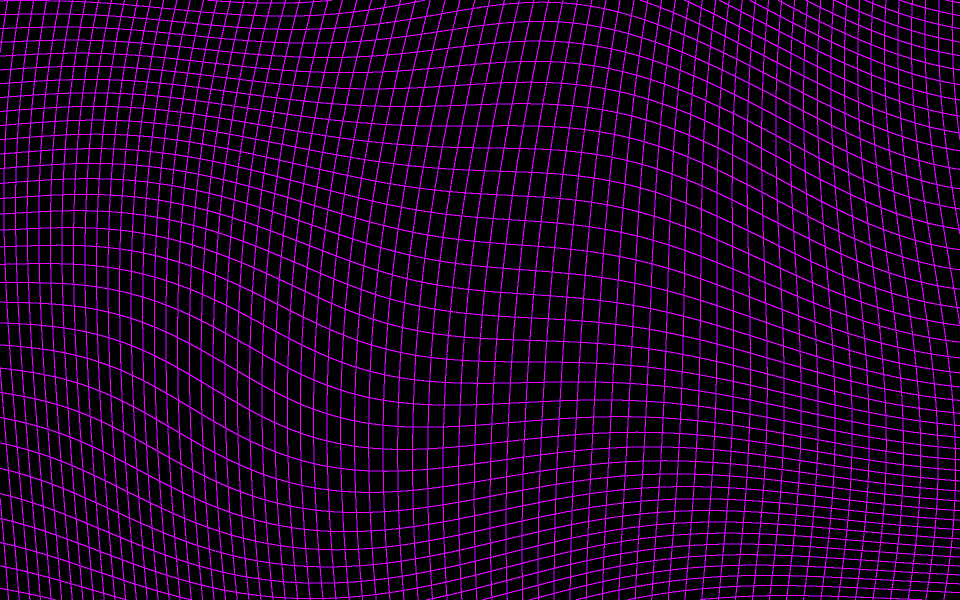}
  \caption{
  Various animated effects. Each pair is showing the effect with our technique with and without enabling the field $d$ (without the bands do not adapt). Please refer to text for details and Table~\ref{tbl:links} for links to shadertoy implementations. \label{fig:anims}
  }
\end{figure*}

\newpage

\balance
\bibliographystyle{ACM-Reference-Format}

\bibliography{biblio}       

\end{document}